\documentclass[a4paper,twoside,prd,showpacs,nofootinbib,preprintnumbers,twocolumn]{revtex4}
\usepackage{amssymb,amsmath,bm,natbib}
\usepackage{color}
\usepackage{slashed}
\usepackage{graphics}
\usepackage{graphicx}
\usepackage[utf8]{inputenc}
\usepackage[caption=false]{subfig}
\usepackage{hyperref}
\usepackage{url}
\usepackage{dsfont}
\usepackage{float} 
\usepackage{cancel}

\newcommand{\epsp}{\mbox{$\epsilon^\prime/\epsilon$}}
\newcommand{\eq}[1]{Eq.~\eqref{#1}}

\newcommand{\e}{\ensuremath{\mathrm{e}}}
\renewcommand{\i}{\ensuremath{\mathrm{i}}}

\begin{document}
\preprint{PSI-PR-19-20,   UZ-TH  44/19, INT-PUB-19-047}
	\title{$Z^\prime$ models with less-minimal flavour violation}
		
\author{Lorenzo Calibbi}
\email{calibbi@nankai.edu.cn}
\affiliation{School of Physics, Nankai University, Tianjin 300071, China}
	
\author{Andreas Crivellin}
\email{andreas.crivellin@cern.ch}
\affiliation{Paul Scherrer Institut, CH--5232 Villigen PSI, Switzerland}
\affiliation{Physik-Institut, Universit\"at Z\"urich, Winterthurerstrasse 190, CH--8057 Z\"urich, Switzerland}

\author{Fiona Kirk}
\email{fiona.kirk@psi.ch}
\affiliation{Paul Scherrer Institut, CH--5232 Villigen PSI, Switzerland}
\affiliation{Physik-Institut, Universit\"at Z\"urich, Winterthurerstrasse 190, CH--8057 Z\"urich, Switzerland}
		
\author{Claudio Andrea Manzari}
\email{claudioandrea.manzari@physik.uzh.ch}
\affiliation{Paul Scherrer Institut, CH--5232 Villigen PSI, Switzerland}
\affiliation{Physik-Institut, Universit\"at Z\"urich, Winterthurerstrasse 190, CH--8057 Z\"urich, Switzerland}

\author{Leonardo Vernazza}
\email{lvernazz@staffmail.ed.ac.uk}
\affiliation{Nikhef, Science Park 105, NL-1098 XG Amsterdam, The Netherlands}	
	
\begin{abstract}
We study the phenomenology of simplified $Z^\prime$ models with a global $U(2)^3$ flavour symmetry in the quark sector, broken solely by the Standard Model Yukawa couplings. This flavour symmetry, known as less-minimal flavour violation, protects $\Delta F=2$ processes from dangerously large new physics (NP) effects, and at the same time provides a free complex phase in $b\to s$ transitions, allowing for an explanation of the hints for additional direct CP violation in kaon decays ($\epsilon^\prime/\epsilon$) and in hadronic $B$-decays ($B\to K\pi$ puzzle). Furthermore, once the couplings of the $Z^\prime$ boson to the leptons are included, it is possible to address the intriguing hints for NP (above the 5$\,\sigma$ level) in $b\to s \ell^+\ell^-$ transitions. Taking into account all flavour observables in a global fit, we find that $\epsilon^\prime/\epsilon$, the $B\to K\pi$ puzzle and $b\to s \ell^+\ell^-$ data can be explained simultaneously. Sizeable CP violation in $b\to s \ell^+\ell^-$ observables, in particular $A_8$, is predicted, which can be tested in the near future, and an explanation of the $B\to K\pi$ and $\epsilon^\prime/\epsilon$ puzzles leads to effects in di-jet tails at the LHC, that are not far below the current limits. Once $b\to s \ell^+\ell^-$ is included, cancellations in di-muon tails, possibly by a second $Z^\prime$, are required by LHC data. \end{abstract}
\maketitle

\newpage
\section{Introduction}

The Standard Model (SM) of particle physics has been very successfully tested with great precision in the last decades. However, it is well known that it cannot be the ultimate theory describing the fundamental constituents and interactions of matter. For example, in order to generate the matter anti-matter asymmetry of the universe, the Sakharov criteria~\cite{Sakharov:1967dj} must be satisfied, one of which is the presence of CP violation. Since the amount of CP violation within the SM is far too small to achieve the observed matter anti-matter asymmetry~\cite{Cohen:1993nk,Gavela:1993ts,Huet:1994jb,Gavela:1994ds,Gavela:1994dt,Riotto:1999yt}, physics beyond the SM with additional sources of CP violation is required. New sources of CP violation could also reconcile the theory prediction~\cite{Buras:2015xba,Buras:2015yba,Bai:2015nea,Kitahara:2016nld,Aebischer:2019mtr}\footnote{These predictions are based on lattice- and dual QCD. Calculations using chiral perturbation theory~\cite{Pallante:2001he,Pich:2004ee,Cirigliano:2011ny,Gisbert:2018tuf} are consistent with the experimental value, but have large errors.} for direct CP violation in kaon decays (\epsp) with the experimental measurements~\cite{Batley:2002gn,AlaviHarati:2002ye,Abouzaid:2010ny}. Similarly, the long-standing ``$B\to K\pi$ puzzle''~\cite{Gronau:1998ep,Buras:2004ub,Baek:2007yy,Fleischer:2007mq}, whose tension~\cite{Fleischer:2017vrb,Fleischer:2018bld} was recently increased by LHCb data~\cite{Aaij:2018tfw}, can be explained~\cite{Crivellin:2019isj}.

It has been shown that models with an additional neutral gauge boson, so-called $Z^\prime$ models, not only explain $\epsilon^\prime/\epsilon$~\cite{Buras:2015yca,Buras:2015jaq,Buras:2015kwd,Buras:2016dxz,Endo:2016tnu,Bobeth:2016llm}, but also provide a promising solution to the $B\to K\pi$ puzzle~\cite{Barger:2009qs,Barger:2009eq}, since they affect electroweak penguin operators~\cite{Fleischer:2008wb,Baek:2009pa}. Furthermore, the anomalies in $b\to s\ell^+\ell^-$ data~\cite{Aaij:2014pli,Aaij:2014ora,Aaij:2015esa,Aaij:2015oid,Khachatryan:2015isa,ATLAS:2017dlm,CMS:2017ivg,Aaij:2017vbb}, which, using a global fit, convincingly point towards NP~\cite{Capdevila:2017bsm, Altmannshofer:2017yso, DAmico:2017mtc, Ciuchini:2017mik, Hiller:2017bzc,Geng:2017svp,Hurth:2017hxg,Alguero:2019ptt,Aebischer:2019mlg,Ciuchini:2019usw}, can be explained within $Z^\prime$ models~\cite{Buras:2013qja,Gauld:2013qba,Gauld:2013qja,Altmannshofer:2014cfa,Crivellin:2015lwa,Falkowski:2015zwa,Celis:2015eqs,Celis:2015ara,Crivellin:2015era,Falkowski:2018dsl,Faisel:2017glo,King:2017anf,Chiang:2017hlj,Crivellin:2017ecl,DiChiara:2017cjq,Ko:2017lzd,Benavides:2018rgh,Maji:2018gvz,Singirala:2018mio,Allanach:2018lvl,Duan:2018akc,King:2018fcg,Kohda:2018xbc,Dwivedi:2019uqd,Foldenauer:2019vgn,Ko:2019tts,Allanach:2019iiy,Altmannshofer:2019xda}. $Z^\prime$ bosons are thus prime candidates for a common explanation of these anomalies. 

For explaining all three anomalies ($\epsilon^\prime/\epsilon$, $B\to K\pi$ and $b\to s\ell^+\ell^-$), small flavour changing couplings to quarks are required that respect the bounds from $\Delta F=2$ processes. Furthermore, as recently shown in Ref.~\cite{Crivellin:2019isj}, the $d-s$ and $s-b$ couplings of the $Z^\prime$ should, after factoring out CKM elements, be of the same order in a common explanation of $\epsilon^\prime/\epsilon$ and the $B\to K\pi$ puzzle. Both the smallness of the flavour changing couplings, as well as the required scaling of $s\to d$ versus $b\to s$ transitions (including a free phase in the latter), point towards a $U(2)^3$ flavour symmetry in the quark sector~\cite{Barbieri:1995uv,Barbieri:1997tu,Barbieri:2011fc,Barbieri:2011ci,Crivellin:2011fb,Barbieri:2012uh,Barbieri:2012bh,Buras:2012sd}\footnote{Similarly, ``standard" minimal flavour violation~\cite{Chivukula:1987fw,Hall:1990ac,Buras:2000dm} (MFV) is based on $U(3)^3$~\cite{DAmbrosio:2002vsn}, however, $U(3)^3$ is strongly broken to $U(2)^3$ by the large third-generation Yukawa couplings.}, also known as ''less-minimal flavour violation". 

In this paper we will examine $Z^\prime$ models in conjunction with a global $U(2)^3$ flavour symmetry. We will work in a simplified framework which only specifies the charges\footnote{We will refer to charges here, but our approach also applies to effective couplings induced e.g. by vector-like fermions~\cite{Bobeth:2016llm}.} of the SM fermions under the new abelian $U(1)^\prime$ gauge symmetry, but not the symmetry breaking sector. We will not impose anomaly cancellation~\cite{Ellis:2017nrp} either, which can be solved at an arbitrary high scale~\cite{Langacker:2008yv,Alonso:2018bcg,Smolkovic:2019jow}. In order to asses the consistency of our model with LHC searches, we consider the bounds on 4-fermion operators (rather than resonant searches), which are model-independent for heavy $Z^\prime$-bosons, since they only depend on the ratio of coupling over mass.
In fact, given the large mass and width of the $Z^\prime$, we assume the signal of $Z^\prime$ production at the LHC would resemble that of a contact interaction, namely, a modification in the tails of the di-jet and di-lepton distributions.

The article is structured as follows: In the next section we will establish our setup and discuss the relevant observables in more detail. Then we will derive less minimal flavour violation applied to $Z^\prime$ models in Sec.~\ref{less-minimal-FV}, before performing the phenomenological analysis in Sec.~\ref{Pheno}. Finally, we conclude in Sec.~\ref{Conclusion}.

\section{Setup and observables}

Let us first review the relevant observables within a generic $Z^\prime$ model with arbitrary couplings to SM fermions, defined by:
\begin{equation}
\mathcal{L} = \sum\limits_{f = u,d,\ell ,\nu } {{{\bar f}_i}{\gamma ^\mu }\left( {\Gamma _{ij}^{fL}{P_L} + \Gamma _{ij}^{fR}{P_R}} \right){f_j}{Z^\prime_\mu }}\,.
\label{Zprime}
\end{equation}
We denote the mass of the $Z^\prime$ by $M_{Z^\prime}$. As outlined in the introduction, we will assume a simplified setup in which the $Z^\prime$ boson originates from a new $U(1)^\prime$ gauge group with the gauge coupling $g^\prime$ and charges $\mathcal{Q}$, but will not specify the corresponding symmetry breaking mechanism, which is very model-dependent. 

\begin{boldmath}
\subsection{$\epsilon'/\epsilon$}
\end{boldmath}
	
For $\epsilon'/\epsilon$, we follow the conventions of Ref.~\cite{Aebischer:2018quc} and use
\begin{equation}
\mathcal{H}_{\Delta S=1} = - \sum_i\frac{C_i(\mu_{ew})}{(1 \text{TeV})^2}O_i\,.
\end{equation}
In order to achieve a numerically large effect, isospin violation (physics that couples differently to up and down quarks) is necessary~\cite{Branco:1982wp}. Since the left-handed current respects isospin due to $SU(2)_L$ gauge invariance, only the operators
\begin{align}
\begin{split}
& O^q_{VLR} = (\bar{s}^{\alpha}\gamma_{\mu}P_Ld^{\alpha})(\bar{q}^{\beta}\gamma^{\mu}P_Rq^{\beta})\,,
\end{split}
\end{align}
with $q=u,d$ and the colour indices $\alpha$ and $\beta$, are relevant for $Z^\prime$ models. The matching to our model leads to the Wilson coefficient
\begin{align}
C^q_{VLR} = - \Gamma^{dL}_{21}\Gamma^{qR}_{11}\dfrac{1\,{\rm TeV}^2}{M_{Z^\prime}^2}\,,
\end{align}
that contributes to  $\epsilon^\prime/\epsilon$ as follows
\begin{equation}
\left(\frac{\varepsilon^\prime}{\varepsilon}\right)_{BSM} \approx 124 \;\Im[C^d_{VLR}-C^u_{VLR}]\,,
\end{equation}
for a matching scale of 1~TeV~\cite{Aebischer:2018quc}.

The experimental average for $\epsilon'/\epsilon$ of the NA48 \cite{Batley:2002gn} and KTeV \cite{AlaviHarati:2002ye,Abouzaid:2010ny} collaborations,
\begin{equation}
\begin{aligned}
\left({\epsilon'}/{\epsilon}\right)_{\rm exp} = (16.6 \pm 2.3) \times 10^{-4}\,,
\end{aligned}
\end{equation}
lies significantly above the SM prediction
\begin{equation}
\begin{aligned}
\left({\epsilon'}/{\epsilon}\right)_{\rm SM}\approx (1.5 \pm 5.5) \times 10^{-4}\, ,
\end{aligned}
\end{equation}
which is based on lattice QCD results~\cite{Bai:2015nea,Blum:2015ywa} and perturbative NLO calculations \cite{Buras:2015yba,Kitahara:2016nld}.

%

\begin{boldmath}
\subsection{ Hadronic $B$-decays }
\end{boldmath}
	
For hadronic $B$-decays (HBD) involving $b\to s$ transitions we use the effective Hamiltonian
\begin{equation}
\mathcal{H}_{\text{eff}}^{\text{NP}} = -\frac{4G_F}{\sqrt{2}}V_{tb}V_{ts}^* \sum_{q=u,d,s,c} (C_5^qO^q_5+C_6^qO^q_6) + h.c.  \,.
\label{hBdH}
\end{equation}
At tree-level only the Wilson coefficient
\begin{equation}
C^q_{5} = - \dfrac{\sqrt{2}}{4G_FV_{tb}V_{ts}^*}\Gamma^{dL}_{23}\Gamma^{qR}_{11}\dfrac{1}{M_{Z^\prime}^2}\,,
\end{equation}
of the operator
\begin{align}
\begin{split}
O^q_5 &= (\bar{s}\gamma^{\mu}P_Lb)(\bar{q}\gamma_{\mu}P_Rq) \,
\end{split}
\label{hBdHOp}
\end{align}
(with $q=u,d$) is generated. As in the case of \epsp~ , the effect from $q=s,c,b,t$ is numerically very small and can thus be neglected.

For the numerical analysis we will rely on the global fit of Ref.~\cite{Hofer:2010ee}, recently updated in Ref.~\cite{Crivellin:2019isj}.\\

\begin{boldmath}
\subsection{$\Delta F=2$ processes}
\end{boldmath}

For concreteness, we give the formula for kaon mixing, following the conventions of Ref.~\cite{Ciuchini:1998ix}:
\begin{equation}
\mathcal{H}_{\text{eff}}^{\Delta S=2} = \sum_{i=1}^{5}C_iQ_i + \sum_{i=1}^{3}\tilde{C}_i\tilde{Q}_i \,.
\label{K0H}
\end{equation}
The only non-zero Wilson coefficients are
\begin{align}
\begin{aligned}
& C_1(\mu_{Z^\prime}) = \frac{1}{2M_{Z^\prime}^2}\left(\Gamma^{dL}_{12}\right)^2\left(1+\frac{\alpha_s}{4\pi}\frac{11}{3}\right)\,,\\
& C_4(\mu_{Z^\prime}) = -\frac{\alpha_s}{4\pi}\frac{\Gamma^{dL}_{12}\Gamma^{dR}_{12}}{M_{Z^\prime}^2}\,,\\
& C_5(\mu_{Z^\prime}) = -\frac{2}{M_{Z^\prime}^2}\Gamma^{dL}_{12}\Gamma^{dR}_{12}\left(1-\frac{\alpha_s}{4\pi}\frac{1}{6}\right)\,,
\end{aligned}
\label{DeltaF2Matching}
\end{align}
associated to the operators
\begin{align}
\begin{aligned}
Q_1 &= (\bar{d}^{\alpha}\gamma_{\mu}P_{L}s^{\alpha})(\bar{d}^{\beta}\gamma^{\mu}P_{L}s^{\beta})\,,\\
Q_4 &= (\bar{d}^{\alpha}P_{L}s^{\alpha})(\bar{d}^{\beta}P_{R}s^{\beta})\,,\\
Q_5 &= (\bar{d}^{\alpha}P_Ls^{\beta})(\bar{d}^{\beta}P_Rs^{\alpha})\,,\\
\end{aligned}
\label{K0Q}
\end{align}
at the matching scale $\mu_{Z^\prime}\sim M_{Z^\prime}$. The chirality-flipped operator $\tilde Q_1$, and its corresponding Wilson coefficient $\tilde C_1$ are obtained from $Q_1$ and $C_1$ by exchanging $L$ with $R$. In \eq{DeltaF2Matching} we included the matching corrections of Ref.~\cite{Buras:2012fs}, such that the 2-loop renormalization group evolution of Ref.~\cite{Ciuchini:1997bw,Buras:2000if} can be consistently taken into account. For a $Z^\prime$-scale of $5\,$TeV and a low scale of $2\,$GeV (where the bag factors are calculated~\cite{Aoki:2019cca}), we find
\begin{align}
C_1(\mu_{\rm low}) &\approx 0.73C_1(\mu_{Z^\prime})\,,\nonumber\\
C_4(\mu_{\rm low}) &\approx 5.73C_4(\mu_{Z^\prime})+1.66C_5(\mu_{Z^\prime})\,,\;\\
C_5(\mu_{\rm low}) &\approx 0.23C_4(\mu_{Z^\prime}) + 0.87C_5(\mu_{Z^\prime})\,.\nonumber
\end{align}
The analogous expressions for $B_d-\bar B_d$ and $B_s-\bar B_s$ mixing are obtained by obvious changes of indices and slight variations of $\mu_{\rm low}$. For the numerical analysis we use the bag factors given in Ref.~\cite{Bazavov:2016nty}.  

Concerning the experimental bounds, for CP violation in kaon mixing ($\epsilon_K$) we use the value given in Ref.~\cite{Bona:2007vi}
\begin{equation}
0.87 \leq \frac{ \epsilon_K^{\rm SM}+\epsilon_K^{\rm NP}}{\epsilon_K^{SM}} \leq 1.39\,,\quad (95\% {\rm C.L.})\,,
\end{equation}
while for $B_d-\bar B_d$ and $B_s-\bar B_s$ mixing we parametrize the two-dimensional fit  result of Ref.~\cite{Bona:2007vi}.\\


\begin{boldmath}
\subsection{$b\rightarrow s \ell^+\ell^-$}
\end{boldmath}

Defining the Hamiltionian
\begin{equation}
\mathcal{H}_{eff} = -\frac{4G_F}{\sqrt{2}}V_{tb}V_{ts}^* \sum_i(C_iO_i+C_i'O_i')\,,
\end{equation}
with the operators
\begin{align}
\begin{split}
O_9^{(\prime)} &= \frac{e^2}{16\pi^2}(\bar{s}\gamma_{\mu}P_{L(R)}b)(\bar{\ell}\gamma^{\mu}\ell)\,,\\
O_{10}^{(\prime)} &= \frac{e^2}{16\pi^2}(\bar{s}\gamma_{\mu}P_{L(R)} b)(\bar{\ell}\gamma^{\mu}\gamma_5\ell)\,.
\end{split}
\end{align}
we get contributions to the Wilson coefficients
\begin{align}
\begin{aligned}
C_9^{(\prime)} &= -\frac{16\pi^2}{e^2}\frac{\Gamma^{dL(R)}_{23}(\Gamma^{\ell L}_{22}+\Gamma^{\ell R}_{22})}{4 \sqrt{2} G_F  M_{Z^\prime}^2 V_{tb}V_{ts}^*}\,,\\
C_{10}^{(\prime)} &= -\frac{16\pi^2}{e^2}\frac{\Gamma^{dL(R)}_{23}(\Gamma^{\ell R}_{22}-\Gamma^{\ell L}_{22})}{4\sqrt{2}G_F M_{Z^\prime}^2 V_{tb}V_{ts}^*}\,,
\end{aligned}
\end{align}
in the concrete case of $b\rightarrow s\mu^+\mu^-$ transitions.

In the following numerical analysis, we make use of the global fits in Ref.~\cite{Descotes-Genon:2015uva,Capdevila:2017bsm}. For example, in the simplest case of $C_9$ with muons only, one has $-3.04<C_{9\mu}^{NP}<-0.76$.\\


%
%
%
%
%
%
%
%

\begin{boldmath}
\subsection{$Z-Z^\prime$ mixing}
\end{boldmath}
\label{Z-Zp}

The $Z^\prime$ boson can mix with the SM $Z$, modifying the couplings of the latter to fermions~\cite{Babu:1996vt,Babu:1997st,Langacker:2008yv,Buras:2014yna}. There is no symmetry which can prevent this mixing, and even if it should vanish at a specific scale, it is generated at a different scale via loop effects.

In analogy with Eq.~(\ref{Zprime}), we write the $Z$ couplings as
\begin{equation}
\mathcal{L}_{Z} = \sum\limits_{f = u,d,\ell ,\nu } {{{\bar f}_i}{\gamma ^\mu }\left( {\Delta _{ij}^{fL}{P_L} + \Delta _{ij}^{fR}{P_R}} \right){f_j}{Z_\mu }}\,, 
\label{Zcoup}
\end{equation}
with
\begin{equation}
\Delta _{ij}^{fL,R} = \sin \theta \;\Gamma _{ij}^{fL,R} + \cos \theta \;\Delta _{\rm SM}^{fL,R}{\delta _{ij}}
\end{equation}
where $\theta$ is the $Z-Z^\prime$ mixing angle and $\Delta _{\rm SM}^{fL,R}$ are the couplings within the SM, given by
\begin{equation}
\begin{aligned}
\Delta _{\rm SM}^{dL} &= \frac{{{g_2}}}{{2{c_W}}}\left( {1 - \frac{2}{3}s_W^2} \right),\;\;\Delta _{\rm SM}^{dR} =  - \frac{{{g_2}s_W^2}}{{3{c_W}}}\,,\\
\Delta _{\rm SM}^{uL} &= \frac{{ - {g_2}}}{{2{c_W}}}\left( {1 - \frac{4}{3}s_W^2} \right),\;\;\Delta _{\rm SM}^{uR} = \frac{{2{g_2}}}{{3{c_W}}}s_W^2\,,\\
\Delta _{\rm SM}^{\ell L} &= \frac{{{g_2}}}{{2{c_W}}}\left( {1 - 2s_W^2} \right),\;\;\Delta _{\rm SM}^{\ell R} =  - \frac{{{g_2}s_W^2}}{{{c_W}}}\,,\\
\Delta _{\rm SM}^{\nu L} &= \frac{{ - {g_2}}}{{2{c_W}}}\,.
\end{aligned}
\end{equation}
The contributions to flavour processes are obtained from the expressions in the previous subsections by replacing $\Gamma$ with $\Delta$ and $Z^\prime$ with $Z$. Note that the contribution of $Z-Z^\prime$ mixing to $\Delta F=2$ processes is suppressed by $\sin^2\theta$, while its contribution to other observables involves only $\sin\theta$. Therefore, the effect of $Z-Z^\prime$ mixing in $\Delta F=2$ processes can be neglected.

Turning to $Z$ couplings to fermions, the best bounds on quark couplings come from $Z\to b\bar b$. Here, due to the forward-backward asymmetry, there is a slight preference for NP effects related to right-handed bottom quarks~\cite{deBlas:2016nqo}:
\begin{equation}
\begin{aligned}
\Delta_{33}^{dR}-\Delta _{\rm SM}^{dR}&=0.012\pm0.004\,,\\
\Delta_{33}^{dL}-\Delta _{\rm SM}^{dL}&=0.0015\pm0.0007\,.
\end{aligned}
\label{ZZpbb}
\end{equation}
If the $Z^\prime$ couples also to leptons, the bounds from $Z\to\ell^+\ell^-$ are very stringent~\cite{Erler:2009jh}. One can estimate the effect to be at most around
0.2\%~\cite{Tanabashi:2018oca}. More concretely, for vectorial couplings to muons and electrons one has
\begin{equation}
\begin{aligned}
-0.0034<\Delta_{22}^{\ell}-\Delta _{SM}^{\ell}<0.0031\,,\\
0.0001<\Delta_{11}^{\ell}-\Delta _{SM}^{\ell}<0.0016\,,
\end{aligned}
\end{equation}
with $\Delta_{ij}=\Delta^L_{ij}+\Delta^R_{ij}$. In addition, there are stringent bounds from from $Z\to\nu\nu$
\begin{equation}
2.9676<\sum_{i,j=1}^3\left|\dfrac{\Delta_{ij}^{\nu}}{\Delta _{SM}^{\nu}}\right|^2<3.0004\, ,
\end{equation}
however, since the measurement does not distinguish between the neutrino flavours, this bound can always be avoided by adjusting the charges of the tau leptons.\\

\subsection{LHC searches}

In our phenomenological analysis we will consider a heavy $Z^\prime$ boson, whose width is very large (of the order of its mass). As a consequence, bounds from searches for narrow resonances can not be directly applied. Although the state might be produced on-shell, as an effect of the large width, the signal mimics that of a contact interaction, i.e.~a change in the tail of di-jet and di-lepton distributions, as we checked by means of a simulation. In addition, it is always possible to rescale the mass and the coupling constant by the same factor, leaving the predictions for flavour observables invariant (despite small logarithmic corrections). To a good approximation, one can thus use the bounds on 4-fermion operators from lepton or jet tails which only depend on the ratio of couplings (times charges) squared divided by the mass squared. Concerning 4-quark operators the current bounds are between $(0.15/{\rm TeV})^2$ and $(0.3/{\rm TeV})^2$~\cite{Sirunyan:2017ygf}. For 2-quark-2-lepton operators the bounds on the Wilson coefficient (without any normalization factor in the effective Lagrangian) with muons are between $(0.12/{\rm TeV})^2$ and $(0.18/{\rm TeV})^2$~\cite{Aaboud:2017buh}. Here both analyses assume quark flavour universality.

\subsection{Landau pole}

In presence of sizeable $U(1)^\prime$ charges, the renormalisation group (RG) running of the gauge coupling $g^\prime$ may generate a Landau pole at unacceptably low energies. This sets an additional constraint to our model, that we study considering the 1-loop RG equation
\begin{equation}
\frac{1}{\alpha^\prime}(\mu)=\frac{1}{\alpha^\prime}(\bar{\mu})-\frac{b}{2\pi}\log(\mu/\bar{\mu})\,,
\end{equation}
where $\alpha^\prime\equiv g^{\prime2}/4\pi$, $\bar{\mu}$ is a low-energy scale, and $b$, the coefficient of the $\beta$-function, is given in terms of the $U(1)^\prime$ charges of the SM fermions by
\begin{align}
\nonumber
b =\frac{2}{3}\sum_{i=1}^3\bigg[&6 \mathcal{Q}_{Q_i}^2+
3\left(\mathcal{Q}_{u_i}^2+\mathcal{Q}_{d_i}^2\right) + 2\mathcal{Q}_{L_i}^2+\mathcal{Q}_{e_i}^2\bigg] \,.
\end{align}
We define the Landau pole scale $\mu_{\rm LP}$ as the scale at which the gauge coupling diverges, which at 1 loop is given by
\begin{equation}
\mu_{\rm LP} = \bar{\mu} \, \exp\left({\frac{2\pi}{b\, \alpha^\prime(\bar{\mu})}}\right)\,.
\end{equation}

\begin{boldmath}
\section{$U(2)^3$-flavour}
\label{less-minimal-FV}
\end{boldmath}

Since only the third-generation-Yukawa couplings are sizeable, the quark sector of the SM Lagrangian possesses an approximate global $U(2)^3=U(2)_Q\times U(2)_u\times U(2)_d$ flavour symmetry for the first two generations of quarks. Here $Q$ and $u$ and ($d$) refer to the left-handed quark $SU(2)_L$ doublet and the right-handed  up (down) quark $SU(2)_L$ singlet, respectively (see Table~\ref{U2fReps})). We assume that this $U(2)^3$-symmetry is respected by the gauge sector and is only broken by the SM Yukawa couplings (which in turn arise from the unspecified $U(1)^\prime$-breaking sector). Therefore, the $U(1)^\prime$-charges must be equal for the first two generations, leading to the following $U(1)^\prime$-charge matrices in flavour space (i.e. in the interaction basis):
\begin{align}
\mathcal{Q}_Q= &\text{diag}(\mathcal{Q}_{Q_{12}},\mathcal{Q}_{Q_{12}},\mathcal{Q}_{Q_3})\,,\nonumber\\
\mathcal{Q}_u= &\text{diag}(\mathcal{Q}_{u_{12}},\mathcal{Q}_{u_{12}},\mathcal{Q}_{u_3})\,,\nonumber\\
\mathcal{Q}_d=&\text{diag}(\mathcal{Q}_{d_{12}},\mathcal{Q}_{d_{12}},\mathcal{Q}_{d_3})\,.\label{FNcharges}
\end{align}

\begin{table}[t]
\centering
\begin{tabular}{|c|c|c|c|}
\hline
 & $U(2)_Q$ &  $U(2)_u$ &  $U(2)_d$ \\
 \hline
 \hline
$\left(Q_1, Q_2\right)$ &  2 & 1 & 1 \\
$\left(u_1, u_2\right)$ &  1 & 2 & 1 \\
$\left(d_1, d_2\right)$ &  1 & 1 & 2 \\
\hline
$Q_3,u_3,\,d_3$ &  1 & 1 & 1\\
\hline
\hline
$\Delta_u$ & 2 & $\bar{2}$ & 1\\
$\Delta_d$ & 2 & 1 & $\bar{2}$\\
\hline
$X_t$ & 2 & 1 & 1\\
$X_b$ & 2 & 1 & 1\\
\hline
\end{tabular}
\caption{$U(2)^3$-representations of the quark fields and spurions in our model.\label{U2fReps}}
\end{table}

In order to recover the small quark masses of the first two generation quarks, as well as the suppressed off-diagonal elements of the CKM-matrix, the $U(2)^3$ symmetry must be broken. Following the strategy presented in Ref.~\cite{Barbieri:1995uv,Barbieri:2011ci,Barbieri:2012uh,Barbieri:2012bh}, the Yukawa couplings of the Lagrangian
\begin{equation}
\mathcal{L}_Y=Q_iY^d_{ij} d_jH + Q_iY^u_{ij} u_j\tilde{H}\,,\, +{\rm h.c. }\,.
\end{equation}
can be written as
\begin{align}
\begin{split}
\dfrac{Y^u}{y_t}=
\begin{pmatrix}
\Delta _u
&	\vline
& X_{t}\\
\hline
\begin{matrix}
0 & 0
\end{matrix}
&	\vline
& 	1
\end{pmatrix}\,,\,
\dfrac{Y^d}{	y_b}=
\begin{pmatrix}\Delta _d
&	\vline
& X_{b}\\
\hline
\begin{matrix}
0 & 0
\end{matrix}
&	\vline
& 	1
\end{pmatrix}\,.
\end{split}\label{YukawaSpurions}
\end{align}
Here $y_{t,b}=\frac{m_{t,b}}{v}$ ($v\approx 174 \,\text{GeV}$) are the Yukawa couplings of the third generation quarks. The minimal spurion sector consisting of $\Delta_{u,d}$ and $X_{t,b}$ is given in Table~\ref{U2fReps}. Using $U(2)$ transformations, the spurions $\Delta_{u,d}$ and $X_{t,b}$ can, without loss of generality, be written as
\begin{align}
\begin{aligned}
\Delta _u &= U^u \text{diag}(\lambda_u,\lambda_c )\,,\; X_t=x_t\e^{\i\phi_t}
\begin{pmatrix}
0\, \\1\,
\end{pmatrix},\\
\Delta _d &= U^d \text{diag}(\lambda_d,\lambda_s )\,,\; X_b=x_b\e^{\i\phi_b}
\begin{pmatrix}
0\, \\1\,
\end{pmatrix},
 \end{aligned}
 \label{DeltaYudDecomp}
\end{align}
where $U^u$ and $U^d$ are unitary $2\times 2$ matrices. The parameters
\begin{align}
\begin{aligned}
\lambda_u\approx \frac{m_u}{m_t}\,, \; \lambda_c\approx \frac{m_c}{m_t}\,,\;
\lambda_d\approx \frac{m_d}{m_b}\,, \;\lambda_s\approx \frac{m_s}{m_b}\,,
\end{aligned}
\label{lambdas}
\end{align}
are $\mathcal{O}(|V_{cb}|\approx 4\times 10^{-2})$ and control the $U(2)^3$-breaking. 

In order to arrive at the mass basis, we diagonalise $Y^u$ and $Y^d$ as follows
\begin{align}
\begin{aligned}
V^{u\dagger} Y^u W^u &= \text{diag}(y_u,y_c,y_t)\,,\\
V^{d\dagger} Y^d W^d &= \text{diag}(y_d,y_s,y_b)\,,
 \end{aligned}
\label{YDiag}
\end{align}
where $V^{u,d}$ ($W^{u,d}$) are unitary the matrices transforming the left- (right-) handed up- and down-type fields. These matrices can be obtained by diagonalizing $Y^qY^{q\dagger}$ ($Y^{q\dagger}Y^q$) in three steps, such that they take the form
\begin{align*}
V^d=R_{12}(\theta_{ds},\phi _{ds})\times R_{23}(\theta _{sb},\phi _{sb})\times R_{13}(\theta _{db},\phi _{db}),
\end{align*}
(and equivalent for $V^u$) as a product of three rotations. Here, $R_{ij}$ is the unitary matrix describing the mixing in the $ij$-sector. $R_{12}$, for example, is of the form
\begin{align}
\begin{split}
R_{12}\left(\theta,\phi\right)= &\begin{pmatrix}
{\rm cos}(\theta) & e^{i\phi}{\rm sin}(\theta) & 0\\
-e^{-i\phi}{\rm sin}(\theta) & {\rm cos}(\theta) & 0\\
0 & 0 & 1
\end{pmatrix}\,.
\end{split}
\label{RotMat}
\end{align}
In order to determine $V^{u,d}$, we first choose an angle $\theta_{ds,uc}$ and a phase $\phi _{ds,uc}$ such that the matrices $U^{u,d}$ in Eq.~(\ref{YukawaSpurions}) are eliminated. Subsequently, we perform a perturbatively diagonalization of the $23$- and the $13$-sector. Keeping only leading-order terms, we obtain
\begin{align}
\begin{aligned}
V^u=\, &R_{12}(\theta_{uc},\alpha_u) \times R_{23}(x_t c_{uc},\phi_t)\\
&\times R_{31}(x_t s_{uc},-(\alpha_u+\phi_t))\,,\label{Vu}\\
V^d=\, &R_{12}(\theta_{ds},\alpha_d) \times R_{23}(x_b c_{ds},\phi_b)\\
&\times R_{31}(x_b s_{ds},-(\alpha_d+\phi_b))\,,
\end{aligned}
\end{align}
where $c_{ab}=\cos (\theta _{ab})$ and $s_{ab}=\sin (\theta _{ab})$. Explicitly, $V^d$ is given by
\begin{equation}
V^d=\begin{pmatrix}
c_{ds} & e^{i \alpha _d} s_{ds} & 0\\
-e^{-i \alpha _d}s_{ds} & c_{ds} & e^{i\phi _b}x_b\\
e^{-i(\alpha _d +\phi _b)}x_b s_{ds} & -e^{-i\phi _b}x_b c_{ds} & 1
\end{pmatrix}.
\end{equation}
Despite our minimal choice of spurions, there is still a flavour mixing between right-handed fields. However, this effect is suppressed by the parameters $\lambda$ in Eq.~(\ref{lambdas}) with respect to the mixing of the left-handed fields. Neglecting the first generation couplings $\lambda_{u,d}$, we obtain
\begin{equation}
W^d = \begin{pmatrix}
1 & 0 & 0 \\
0 & 1 & \lambda_s\cos (\theta_{ds}) e^{i\phi _b}\\
0 & -\lambda_s\cos (\theta_{ds}) e^{-i\phi _b} & 1
\end{pmatrix}\,,
\label{Wd}
\end{equation}
and a similar expression for $W^u$. 

Now we can determine the the $Z'$-couplings to the quarks which are given by
\begin{equation}
\begin{split}
&\Gamma ^{uL}\equiv g^\prime V^{u\dagger}\mathcal{Q}_Q V^u\,, \qquad \Gamma ^{uR}\equiv g^\prime W^{u\dagger}\mathcal{Q}_u W^u\,, \\
&\Gamma ^{dL}\equiv g^\prime V^{d\dagger}\mathcal{Q}_Q V^d\,, \qquad \Gamma ^{dR}\equiv g^\prime W^{d\dagger}\mathcal{Q}_d W^d\,.
\end{split}
\label{Gammas}
\end{equation}
Making use of the unitarity of the matrices $V^{u,d}$ and $W^{u,d}$, and comparing the results with the elements of the CKM matrix, defined by $V=V^{u\dagger}V^d$, we obtain
\begin{align}
\begin{aligned}
\Gamma _{12}^{dL}
=\,&g^\prime c_K X_Q V_{td}^*V_{ts}\, ,\\
\Gamma _{13}^{dL}
=\,&g^\prime c_B e^{i\alpha _B}X_QV_{td}^* V_{tb}\, ,\\
\Gamma _{23}^{dL}
=\,&g^\prime c_B e^{i\alpha _B}X_QV_{ts}^* V_{tb}\, ,\\
\Gamma _{11}^{qR} =\,& g^\prime \mathcal{Q}_{q_{1,2}}, \quad q=u,d\,,\\
\Gamma _{23}^{dR}=\,&-g^\prime X_d\lambda_s x_b e^{\i\phi _b} \cos (\theta_{ds})\,,
\end{aligned}
\end{align}
at leading order in our perturbative diagonalization. Here we have introduced the notation
\begin{align}
\begin{split}
X_Q &= ( \mathcal{Q}_{Q_3}-\mathcal{Q}_{Q_{1,2}})\,,\\
X_d &= (\mathcal{Q}_{d_3}-\mathcal{Q}_{d_{1,2}})\,,\\
X_{ud} &= (\mathcal{Q}_{u_{1,2}}-\mathcal{Q}_{d_{1,2}})\,,
\end{split}
\label{Qabb}
\end{align}
and the order-one parameters
\begin{align}
c_B&=\frac{x_b}{|e^{-i\phi _t}x_b-e^{-i\phi _b} x_t|},\quad c_K=c_B^2\,,\label{cKcB}
\end{align}
together with the free phase 
\begin{equation}
\alpha _B=\phi_b+\text{arg}(e^{-i\phi _t}x_b-e^{-i\phi _b} x_t).\label{alphaB}
\end{equation}
Note that in the limit $x_t,\phi_t\to 0$ (as in Ref.~\cite{Crivellin:2015lwa}) $c_B\to 1$ and $\alpha_B\to \pi$.

\begin{figure}[t]
	\includegraphics[width=0.5\textwidth]{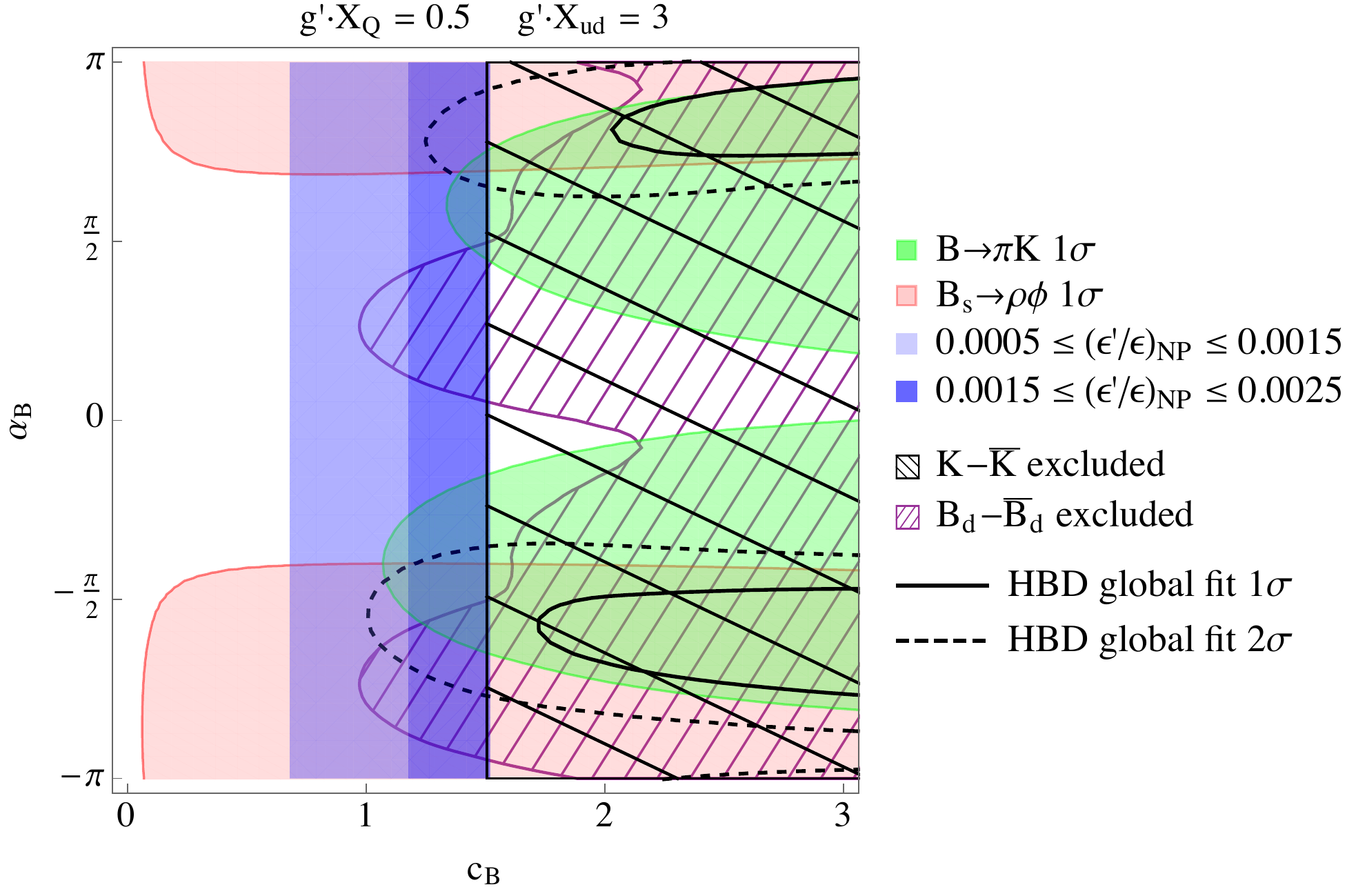}
	\caption{\label{fig:HBD} Preferred regions in the $c_B-\alpha_B$ plane from $B\rightarrow \pi K$ and $B_s\rightarrow \rho\phi$ ($1\sigma$) together with regions from the global fit, including all observables on hadronic $B$-decays ($1\sigma$ and $2\sigma $) as well as $\epsilon^{\prime}/\epsilon$ for $g^{\prime}X_Q=0.5$ and $g^{\prime}X_{ud}=3$. Here we marginalized over $\mathcal{Q}_{u_{1}}+\mathcal{Q}_{d_{1}}$.}
\end{figure}

The $U(2)$ flavour symmetry can be extended to the lepton sector, resulting in a global $U(2)^5$ symmetry. However, the $U(2)$-breaking pattern in the lepton sector cannot be obtained from the PMNS-matrix in the same way as it is obtained from the CKM-matrix in the quark sector, this due to the probable presence of right-handed neutrinos in the see-saw mechanism. Furthermore, other flavour symmetries~\cite{Araki:2012ip}, such as $L_\mu-L_\tau$~\cite{Binetruy:1996cs,Bell:2000vh,Choubey:2004hn,Dutta:1994dx,Heeck:2011wj}, can generate the correct structure for the PMNS-matrix. In what follows, we will consider two scenarios: scenario LFU, the scenario of lepton flavour universality (LFU), which corresponds to a $U(3)^2$-symmetry in the lepton sector, and scenario $L_\mu-L_\tau$, the scenario of lepton flavour universality violation in the form of a $L_\mu-L_\tau$-symmetry. For the calculation of our observables we will nonetheless use generic, but flavour diagonal, couplings of the $Z^\prime$ to leptons
\begin{align}
\begin{aligned}
\Gamma _{ij}^{\ell L}&=g'\mathcal{Q}_{L_{i}}\delta_{ij}\,,
\Gamma _{ij}^{\ell R}&=g'\mathcal{Q}_{e_{i}}\delta_{ij}\,.
\end{aligned}
\end{align}
Note that the LFU scenario automatically prevents dangerous $Z^\prime$-mediated contributions to lepton-flavour-violating processes (cf.~\cite{Calibbi:2017uvl} for a recent review), while in the case of $L_\mu-L_\tau$ we have to assume that the charged lepton Yukawa matrix is (quasi-)diagonal in the interaction basis, a situation that can arise in presence of an additional, possibly discrete, flavour symmetry.

\section{Phenomenological Analysis}\label{Pheno}

\begin{figure*}
	\includegraphics[width=1\textwidth]{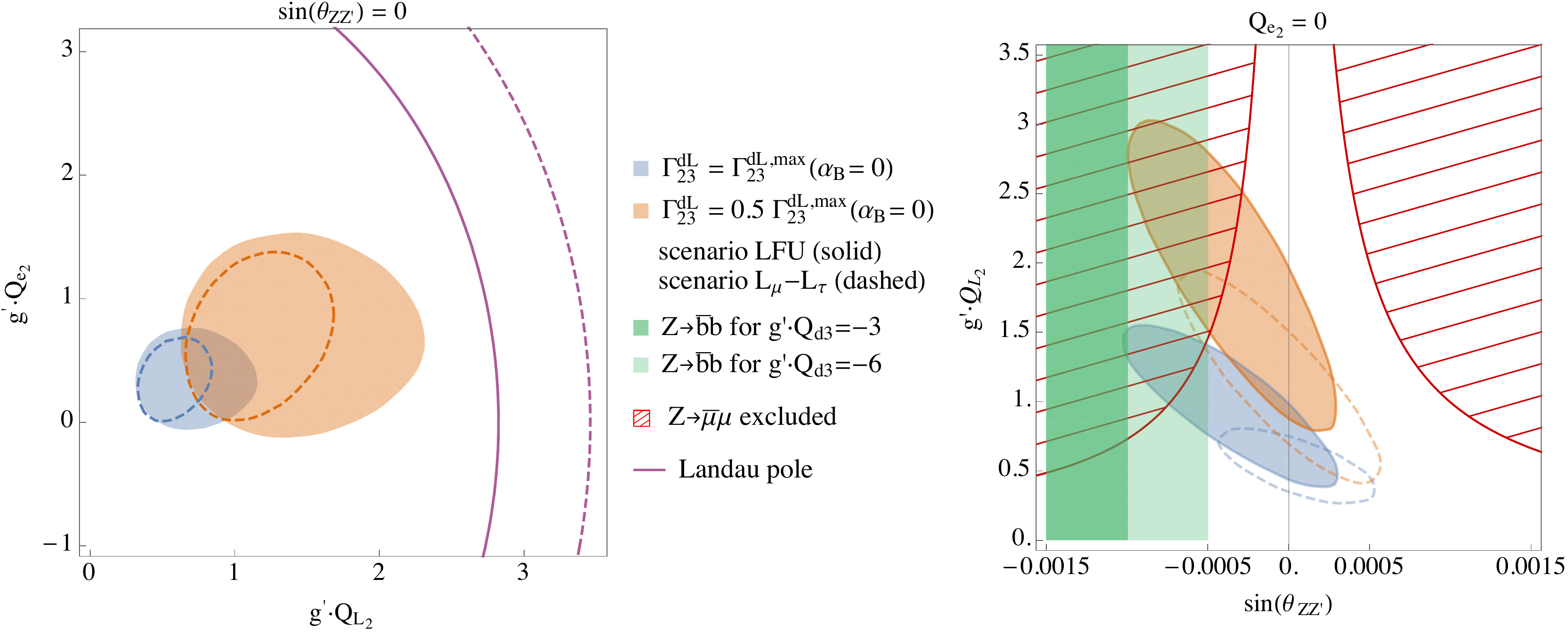}
	\caption{\label{fig:bsllZZp} Left: Preferred regions from $b\rightarrow s\ell^+ \ell^-$ data for different values of $\Gamma^{dL}_{23}$ assuming no $Z-Z^\prime$ mixing for $m_{Z^\prime=5\,}$TeV. The filled regions refer to case 1) with LFU while the regions within the dashed curves correspond to the $L_\mu-L_\tau$ scenario. The corresponding regions with a Landau pole above $50\,$TeV lie to the left of the purple lines. Right: Preferred regions in the $g^{\prime}Q_{L_2}$-$\text{sin}(\theta_{ZZ^{\prime}})$ plane from $b\rightarrow s\ell^+\ell^-$,  $Z\rightarrow \bar{b}b$ and $Z\rightarrow \bar{\mu}\mu$ with $Q_{e_2} = 0$ and $m_{Z^\prime=5\,}$TeV. Again, solid (dashed) lines correspond to the LFU ($L_\mu-L_\tau$) scenario. }
\end{figure*}

In a first step we look at the quark sector only. Among the $\Delta F=2$ processes, we have effects in $K-\bar{K}$, $B_s-\bar{B}_s$ and $B_d-\bar{B}_d$ mixing. Due to the $U(2)^3$ flavour symmetry, the bounds from $D^0-\bar D^0$ are always sub-leading compared to those from $K-\bar{K}$ mixing, where the phase fixed to ${\rm Arg}[(V_{ts}V_{td}^*)^2]$ leads to stringent bounds from $\epsilon_K$. This leads to a maximally allowed value (at 95\% CL) for the coupling $\Gamma^{dL}_{12}$ of
\begin{equation}
|g^\prime c_B^2 X_Q| \lesssim 1.1 \frac{M_{Z^\prime}}{5\,{\rm TeV}}=\Gamma^{dL,{\rm max}}_{12}\,.
\end{equation}
Concerning $B_s-\bar{B}_s$ mixing, we note that the bound can always be avoided by an appropriate choice of $\phi_B$ and $s_{bs}$ since
\begin{align}
\begin{split}
{\langle B_s|\mathcal{H}^{NP}|\bar{B}_s \rangle} \sim   X_\mathcal{Q} + 50 s_{bs}\e^{\i (\alpha_B+\phi_b)}X_d/c_B\,,
\end{split}
\label{eq:BSNP}
\end{align}
for natural values of the parameters involved (since $s_{bs}$ is of order of the $V_{cb}$). Therefore, we are left with the slightly less stringent bounds from $B_d-\bar B_d$ mixing~\cite{Bona:2007vi} which are (to a good approximation) unaffected by right-handed $Z^\prime ds$ couplings. Here we have
\begin{equation}
|g^{\prime}c_BX_\mathcal{Q}|\lesssim [0.5-0.95]=\Gamma^{dL,{\rm max}}_{13}(\alpha_B)\,,
\end{equation}
depending on the specific values of $\alpha_B$. 

Concerning direct CP violation we first include $\epsilon^\prime/\epsilon$ in our analysis. Here, the bounds from $\epsilon_K$ (at 95\% CL) leads to a minimal charge difference $X_{ud}=\mathcal{Q}(u_{1,2})-\mathcal{Q}(d_{1,2})$ necessary to get a NP contribution  $(\epsilon^\prime/\epsilon)_{\rm NP}$ in $\epsilon^\prime/\epsilon$:
\begin{equation}
|g^{\prime} X_{ud}| \gtrsim 1.26 \times \frac{(\epsilon^\prime/\epsilon)_{\rm NP}}{10^{-3}}\,.
\end{equation}
Let us turn to CP violation in hadronic $B$-decays (HBD), in particular in $B_s\rightarrow \rho \phi, \,K\bar{K}$ and in $B\rightarrow \pi K, \, \rho K,  \,\pi K^*, \, \rho K^*$. We find that for $g^{\prime}X_Q=0.5$ and $g^{\prime}X_{ud}=3$, all HBD data can be fitted in the same region of the $c_B-\alpha_B$-plane. This is illustrated in Fig.~\ref{fig:HBD}, where we marginalized over $\mathcal{Q}_{u_{1}}+\mathcal{Q}_{d_{1}}$)~\cite{Crivellin:2019isj}. It is also possible to address $\epsilon^\prime/\epsilon$ and the $B\to K\pi$ puzzle simultaneously without violating bounds from $\Delta F=2$ processes. The resulting charges lead to a naive estimate of an interaction strength for the 4-quark operators of $\approx 0.15{\rm TeV}^2$. This is still consistent with LHC searches, but very close to the current exclusion limits.

We move on to the study of $b\rightarrow s\ell^+\ell^-$ transitions. As outlined in the previous section, we consider a scenario with LFU and a scenario with $L_\mu-L_\tau$. In Fig.~\ref{fig:bsllZZp} (left), we show the regions preferred by $b\to s\ell^+\ell^-$~\cite{Descotes-Genon:2015uva} data for different values of $\Gamma^{dL}_{23}$, together with the predictions for a Landau pole at $50\,$TeV. Here the LHC bounds are be respected if we use a ``minimal" charge assignment in the sense that we allow the third generation of left-handed quarks to have non-zero charges, $\mathcal{Q}_{Q_{3}}$, but set all other quark couplings to zero. This avoids couplings of the $Z^\prime$ boson to the valence quarks of the proton.

So far, we did not consider the effect of $Z-Z^{\prime}$ mixing. In the absence of couplings of the $Z^\prime$-boson to leptons, the most stringent constraints come from $Z\rightarrow \bar{b}b$ ~\cite{deBlas:2016nqo}. However, once the couplings to the leptons are included, $Z\to \bar{\mu}\mu$  gives more stringent bounds~\cite{ALEPH:2005ab}. Furthermore, $Z-b-s$ couplings induced by $Z-Z^{\prime}$ mixing
have an important impact on the global fit of $b\to s\ell^+\ell^-$ data~\cite{Alguero:2019ptt}. This situation is depicted in the plot at the right-hand side of Fig.~\ref{fig:bsllZZp}, where the preferred regions from $b\to s\ell^+\ell^-$ data (determined by FLAVIO~\cite{Straub:2018kue}) and the regions excluded by $Z\rightarrow \bar{\mu}\mu$ are shown in the case of $\alpha_B=0$, for different values of $\Gamma^{dL}_{23}$ and $g^{\prime}Q_{d_3}$. In this figure we also see that the forward-backward asymmetry in $Z\rightarrow \bar{b}b$ (see Eq.~\ref{ZZpbb}) leads to a preference for non-zero mixing. Note that $\text{sin}(\theta_{ZZ^{\prime}})\sim -5\times 10^{-4}$ gives a good fit to data. A value of this order will have an impact on $\epsilon'/\epsilon$ and hadronic $B$-decays of the order of $10\%$, with respect to the $Z^\prime$ contribution.

\subsection{Benchmark Scenario}

Based on the observations discussed above, we now construct a benchmark scenario (along with our two scenarios concerning the lepton couplings) with the aim of addressing $\epsilon^\prime/\epsilon$, hadronic $B$-decays and $b\to s\ell^+\ell^-$ data simultaneously. We choose $g^\prime=0.6$, $M_{Z^\prime}=6\,$TeV and
\begin{align}
\begin{aligned}
\mathcal{Q}_Q&= (0,0,1),\;
\mathcal{Q}_u= (2,2,1),\;\label{BenchmarkFNcharges}
\mathcal{Q}_d= (\!-4,\!-4,0).\\
\mathcal{Q}_L&= (0,-2,2)\;\;({\rm scenario}\,L_\mu-L_\tau)\,,\\
\mathcal{Q}_L&= (-2,-2,-2)\;\;({\rm scenario\;LFU})\,,\\
\mathcal{Q}_e&= (0,0,0)\,,\;\text{sin}(\theta_{ZZ^{\prime}})= 0.001\,.
\end{aligned}
\end{align}
This benchmark point leads to a Landau pole at $\sim 50\,$TeV ($\sim 60\,$)TeV for the LFU ($L_\mu-L_\tau$) scenario. 

The interaction strength of 2-quark-2-muon operators in this scenario is $\approx (0.25{\rm TeV})^2$, which is in conflict with LHC bounds. In order to reconcile the model with LHC data, one could obviously reduce the coupling strength to the right-handed up- and down-quarks, which would decrease the effect in $\epsilon^\prime/\epsilon$, or reduce the coupling strength to muons, which would weaken the impact of our model on $b\to s\ell^+\ell^-$ data. We will pursue another possibility here, making use of the sensitivity to interference of the bounds on 4-fermion operators from LHC-searches in di-lepton or di-jet tails. We suppose the existence of a second neutral gauge boson, $Z^{\prime\prime}$. If the product of the $U(1)^{\prime\prime}$-charges of the right-handed quark and muon has the opposite sign to the equivalent product of $U(1)^{\prime}$-charges (as given in \eq{BenchmarkFNcharges}), destructive interference in LHC searches appears. If we further assume that the $U(1)^{\prime\prime}$-charges of the left-handed quarks respect $U(3)$ flavour symmetry (i.e.~they are equal), only the LHC-searches are affected, while the flavour observables are still governed by $Z^{\prime}$ alone.

\begin{figure*}[t]
	\includegraphics[width=0.7\textwidth]{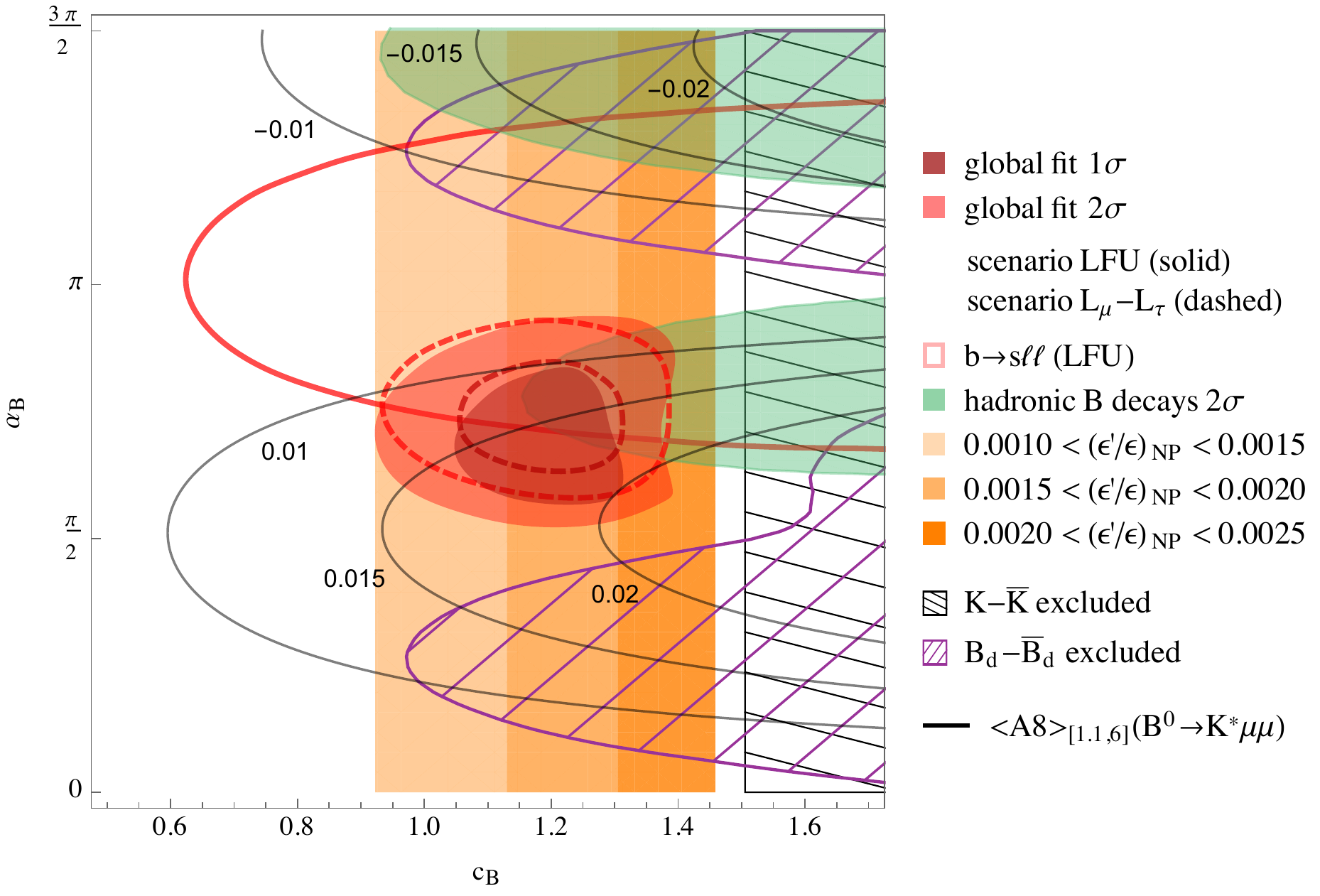}
	\caption{\label{fig:bm} Preferred regions of the combined fit (red) to $\epsilon'/\epsilon$, $b\rightarrow s \ell^+\ell^-$, $K-\bar K$ mixing and $B_d-B_d$ mixing at $1\sigma$ and $2\sigma$ for our benchmark point with the two scenarios LFU and $L_\mu-L_\tau$. In addition, the individual regions from hadronic $B$-decays and $b\rightarrow s \ell^+\ell^-$ data, as well as the regions excluded by $B_d-\bar B_d$ mixing and $\epsilon_K$ are shown and the contour lines for $\text{A}_8(B^0\rightarrow K^*\mu\mu)$ in the $q^2$ interval $\left[1.1, 6\right]$ are depicted.}
\end{figure*}

Now we proceed to the combined analysis of flavour data. Fig.~\ref{fig:bm}$\,$ shows the preferred regions of the combined fit of $b\rightarrow s \ell^+\ell^-$, $\epsilon'/\epsilon$, $\epsilon_K$ and $B_d-\bar B_d$-mixing at $1\sigma$ and $2\sigma$, the preferred/excluded regions of each observable separately, as well as the region preferred by hadronic $B$-decays. We also show the predictions for the $b\rightarrow s \ell^+\ell^-$ observable $\langle\text{A8}(B^0\rightarrow K^*\mu\mu)\rangle_{\left[1.1,6\right]}$~\cite{Aaij:2015oid}, which is especially sensitive to CP violation. A choice of $ \alpha_B\sim\left[2.5 - 3\right]$ and $c_B\sim1.4$ allows us to explain $\epsilon'/\epsilon$, hadronic $B$-decays and $b\rightarrow s \ell^+\ell^-$ data simultaneously at the $2\,\sigma$ level, and to predict $\langle\text{A8}(B^0\rightarrow K^*\mu\mu)\rangle_{\left[1.1,6\right]} \sim\left[ 0.015 - 0.03\right]$, which is agreement with the experimental measurements $\langle\text{A8}\rangle_{\left[1.1,6\right]}^{exp} = -0.047\pm0.058$~\cite{Aaij:2015oid}. With the expected future improvements~\cite{Kou:2018nap,Cerri:2018ypt}, this prediction will soon be testable.

Finally, let us comment on the preliminary results for $K_L\to \pi^0 \nu \bar{\nu}$, where 3 event candidates were observed~\cite{KOTOpresentation,Kitahara:2019lws}. For the best fit point of our LFU($L_\mu-L_\tau$) scenario we obtain a reduction of $\sim-75\%(-30\%)$ with respect to the SM prediction.

\section{Conclusions}\label{Conclusion}

Very interesting deviations from the SM predictions have been found in \epsp, hadronic $B$-decays (HBD) and $b\to s\ell^+\ell^-$ data. In this article we studied these puzzles in a simplified framework involving a heavy $Z^\prime$ boson, but disregarding the explicit form of the symmetry breaking sector. We derived the flavour structure of such models with a $U(2)^3$ symmetry in the quark sector, finding that it is entirely governed by the known CKM elements, as well as two free parameters, a real order-one factor $c_B$ and a complex phase $\phi_B$, which enters $b\to s(d)$ transitions. Importantly, the phase in $s\to d$ transitions is fixed by $V_{tb}V_{ts}^*$, and the corresponding real coefficient $c_K$ is to a good approximation equal to $c_B^2$, making this setup very predictive.

In the phenomenological part of this paper, we first analyzed \epsp~and hadronic $B$-decays (HBD), finding that a common explanation, that respects the bounds from $\Delta_F=2$ processes, is possible. In particular, our setup with less-minimal flavour violation, can always avoid the bounds from $B_s-\bar B_s$ mixing. This cancellation is possible for natural values of the $U(2)^3$ breaking parameters, even then a positive effect in $\epsilon_K$ is predicted. Further, the large isospin violating couplings to quarks required for a common explanation of direct CP-violating in hadronic kaon and $B$ decays, lead to sizeable effects in di-jet tail searches at the LHC, whihc will be testable at the HL-LHC.

Once $b\to s\ell^+\ell^-$ data is included in the analysis, the situation becomes even more interesting. As HBDs require a large phase $\phi_B$, sizeable CP violation in $b\to s\ell^+\ell^-$ observables, in particular in $A_8$, is predicted. We presented a benchmark point, which is capable of providing a common explanation of all flavour data (see Fig.~\ref{fig:bm}). However, the large couplings to up- and down- quarks required by \epsp~and HBD lead to large effects in di-muon tails, excluded by current data. This obstacle can be overcome by postulating destructive interference in LHC searches, e.g. by a second $Z^\prime$ boson with flavour universal couplings to left-handed quarks, such that it does not affect flavour observables. 

In summary, simplified $Z^\prime$ models with less-minimal flavour violation can explain \epsp, HBD and $b\to s\ell^+\ell^-$ data simultaneously, but more new particles are required, as also suggested by the need for a symmetry breaking sector and the presence of a Landau pole at $\approx 50\,$TeV, opening up interesting future directions in model building.

{\it Acknowledgments} --- {\small 
	We thank Emanuele Bagnaschi for very valuable help with LHC simulations and Andrzej Buras for useful discussions. The work of A.C. is supported by a Professorship Grant (PP00P2\_176884) of the Swiss National Science Foundation. L.V. is supported by the D-ITP consortium, a program of NWO funded by the Dutch Ministry of Education, Culture and Science (OCW). A.C. thanks the INT at the University of Washington for its hospitality and the DOE for partial support during the completion of this work.
}

\bibliography{bibliography}

\end{document}